\documentclass[a4paper,12pt]{article}

\usepackage{amsmath,amssymb}
\usepackage[dvips]{graphicx}

\newcommand{\del}{\partial}
\newcommand{\Xdot}{\dot{X}}
\newcommand{\Tbar}{\bar{T}}

\newcommand{\chitilde}{\tilde{\chi}}
\newcommand{\Nbar}{\bar{N}}

\begin{document}
\renewcommand{\thefootnote}{\fnsymbol{footnote}}
\begin{titlepage}
\hfill
{\hfill \begin{flushright} 
HIP-2010-27/TH
\end{flushright}
  }

\vspace*{10mm}

\begin{center}
{\LARGE {\LARGE  
\textbf{Shear viscosity of a highly excited string and the black hole membrane paradigm}}}
\vspace*{18mm}

  {\large Yuya Sasai$^{1,2}$\footnote{sasai@mappi.helsinki.fi}
 and Ali Zahabi$^2$}\footnote{seyedali.zahabi@helsinki.fi}

\vspace*{13mm}
{\large {\it 
$^1$Helsinki Institute of Physics, University of Helsinki, \\
\vspace*{0.5mm} 
P.O.Box 64, FIN-00014 Helsinki, Finland \\
\vspace*{1mm}
$^2$Department of Physics, University of Helsinki, \\
\vspace*{1.5mm}
P.O.Box 64, FIN-00014 Helsinki, Finland
}} \\

\end{center}

\vspace*{16mm}

\begin{abstract}
Black hole membrane paradigm states that a certain viscous membrane seems to be sitting on a stretched horizon of a black hole from the viewpoint of a distant observer. 
We show that the shear viscosity of the fictitious membrane can be reproduced by a highly excited string covering the stretched horizon except for a numerical coefficient.

\end{abstract}

\end{titlepage}

\newpage
\renewcommand{\thefootnote}{\arabic{footnote}}
\setcounter{footnote}{0}

\section{Introduction}
There are several mysteries about black holes which have yet to be solved. One of them is the membrane paradigm \cite{Damour:1978cg, Thorne:1986iy, Parikh:1997ma, Mathur:2010dg}. 
The membrane paradigm states that, in the context of general relativity, transport coefficients such as a viscosity can be obtained on a stretched horizon  of a black hole which is located  near the event horizon. This means that there seems to be a fictitious viscous membrane living on the stretched horizon from the viewpoint of a distant observer. However, the microscopic interpretation of the  membrane  is  not still clear. 

The microscopic description of the fictitious membrane might be understood through string theory which is closely related to black hole physics.
In fact, it has been found that the Bekenstein-Hawking entropy of a  Schwarzshild black hole can be reproduced by a highly excited string covering the stretched horizon except for a numerical coefficient \cite{Susskind:1993ws}.
It has also been proposed that a highly excited string, when we increase the string coupling, becomes a black hole whose horizon radius is of the order of the fundamental string length scale $l_s$ at a critical string coupling \cite{Horowitz:1996nw}. 
That proposal has been examined in the case of BPS states of the heterotic string theory by comparing the logarithm of the number of the BPS states with the area of the stretched horizon of the corresponding BPS black hole \cite{Sen:1995in}. The derivation of the Bekenstein-Hawking entropy from string theory including the  numerically correct coefficient has been found in \cite{Strominger:1996sh, Callan:1996dv}.

So far, the correspondence between string theory and black hole physics has been confirmed mainly by examining their entropies. However, if black holes can be explained by string theory, the membrane paradigm must also be explained by string theory.  Studies related to this problem have been discussed in the context of AdS/CFT correspondence \cite{Kovtun:2003wp, Iqbal:2008by, deBoer:2008gu, Bredberg:2010ky, Faulkner:2010jy}.

Let us assume that the fictitious  membrane on the stretched horizon is composed of a highly excited string according to \cite{Susskind:1993ws}. Then, the viscosity of the membrane must be reproduced by the highly excited string. But what is the meaning of the viscosity of the string? To understand it, 
let us consider a  polymeric liquid as an illustrating example. It is known in polymer physics that the viscosity of a polymeric liquid is in general different from the viscosity of the original solvent because the stress tensor of the polymer itself is added to the stress tensor of the solvent \cite{Doi:1986}. In other words, a polymer possesses its own viscosity. In the same way, a string also possesses its own viscosity because the string has its own stress tensor. In this paper, we calculate the shear viscosity of a highly excited string and find that the shear viscosity of the highly excited string on the stretched horizon  is consistent with the shear viscosity of the fictitious membrane in the membrane paradigm.

This paper is organized as follows. In Sec. \ref{sec:stringhigh}, we introduce the bosonic open string theory. The metric perturbation around the flat background spacetime is considered  to obtain the viscosity of the  string. Then, we review how to evaluate observables in  highly excited string states. In Sec. \ref{sec:viscosstr}, we evaluate the viscosity of the highly excited string by using the Kubo's  formula. The ratio of shear viscosity to entropy density is also found. 
In Sec. \ref{sec:comparison}, we consider a highly excited string on the stretched horizon of the Schwarzshild black hole. By using the results obtained in Sec. \ref{sec:viscosstr}, we  evaluate the viscosity of the highly excited string on the stretched horizon and conclude that the shear viscosity and the ratio of shear viscosity to entropy density agree with the results of the membrane paradigm except for numerical coefficients.
The final section, \ref{sec:summary}, is devoted to the summary and comments.
In the Appendix, we calculate the typical sizes of the highly excited free open string.
  
\section{Open string in highly excited states} \label{sec:stringhigh}
\subsection{Review of bosonic open string theory}
We start with the bosonic string theory in a weakly curved $d+1$ dimensional spacetime, whose metric is given by $g_{\mu\nu}(x)~(\mu, \nu=0,1, \dots d)$ \cite{Green:1987sp}.\footnote{The number of the spatial dimensions is denoted by $d$. We assume $d\ge 3$ because there is no known black hole solution with vanishing cosmological constant when $d<3$.} The world sheet action is 
\begin{align}
S=-\frac{1}{4\pi \alpha'}\int d^2\sigma \sqrt{-\gamma}\gamma^{\alpha\beta}\del_{\alpha}X^{\mu}\del_{\beta}X^{\nu}g_{\mu\nu}(X),
\end{align}
where $\sqrt{\alpha'}\equiv l_s$ is the fundamental string length scale, $\sigma^{\alpha}=(\tau, \sigma) ~(\alpha=0,1)$ are the world sheet coordinates, $X^{\mu}(\sigma)$ are the coordinates of the string in the $d+1$ dimensional spacetime, and   $\gamma_{\alpha\beta}(\sigma)$ is the world sheet metric. Taking the unit gauge,
\begin{align}
\gamma_{\alpha\beta}=\eta_{\alpha\beta},
\end{align}
where $\eta_{\alpha\beta}=(-1,1)$, the action becomes
\begin{align}
S=\frac{1}{4\pi \alpha'}\int d^2\sigma (\Xdot^{\mu}\Xdot^{\nu}-X^{'\mu}X^{'\nu})g_{\mu\nu}(X), \label{eq:actiongauge}
\end{align}
where 
\begin{align}
\Xdot^{\mu}\equiv \frac{\del X^{\mu}}{\del \tau},~~~X^{'\mu}\equiv \frac{\del X^{\mu}}{\del \sigma}.
\end{align}

In order to quantize the string, let us consider the perturbation of the background metric $g_{\mu\nu}(x)$ around the flat spacetime, whose metric is $\eta_{\mu\nu}=(-1, 1, \cdots, 1)$. Then, the action (\ref{eq:actiongauge}) becomes
\begin{align}
S&=S_0+S_1, \\
S_0&=\frac{1}{4\pi \alpha'}\int d^2\sigma~ (\Xdot^{\mu})^2-(X^{'\mu})^2, \\
S_1&=\frac{1}{4\pi \alpha'}\int d^2\sigma~ (\Xdot^{\mu}\Xdot^{\nu}-X^{'\mu}X^{'\nu})h_{\mu\nu}(X), \label{eq:s1}
\end{align}
where $h_{\mu\nu}(X)=g_{\mu\nu}(X)-\eta_{\mu\nu}$ is the perturbation around the flat background metric. $S_0$ is the action for the free string in the flat background spacetime, whose quantization is well known.   We treat  $S_1$ as the perturbation around $S_0$.

For simplicity, we consider the open string. In the case of open string, the mode expansion of $X^{\mu}$ is given by
\begin{align}
X^{\mu}(\tau, \sigma)=\bar{x}^{\mu}+2\alpha' p^{\mu}\tau+i\sqrt{2\alpha'}~ \sum_{\stackrel {\scriptstyle n\neq 0}{\scriptstyle n\in \mathbf{Z}}}
\frac{\alpha_n^{\mu}}{n}e^{-in\tau}\cos (n\sigma), \label{eq:modeexp}
\end{align}
where $0\leq \sigma \leq \pi$, $p^{\mu}$ is the center-of-mass momentum and $X_{cm}^{\mu}\equiv \bar{x}^{\mu}+2\alpha' p^{\mu}\tau$ is the center-of-mass position of the open string at a given $\tau$. 
When the open string is quantized, $\bar{x}^{\mu},~ p^{\mu}$ and the oscillators $\alpha_n^{\mu}$ satisfy the following commutation relations:
\begin{align}
[\bar{x}^{\mu}, p^{\nu}]&=i\eta^{\mu\nu}, \\
[\alpha_m^{\mu}, \alpha_n^{\nu}]&=m\delta_{m+n,0}\eta^{\mu\nu}.
\end{align}

To solve the Virasoro constraints,
\begin{align} 
(\Xdot^{\mu} \pm X^{'\mu})^2=0, \label{eq:virasoro}
\end{align}
we introduce the light-cone coordinates,
\begin{align} 
X^{\pm}\equiv \frac{1}{\sqrt{2}}(X^0\pm X^{d}).
\end{align}
Taking the light-cone gauge, 
\begin{align}
X^{+}=\bar{x}^{+}+2\alpha' p^{+}\tau, \label{eq:lightconegauge}
\end{align}
the Virasoro constraints (\ref{eq:virasoro}) become
\begin{align}
\Xdot^{-}\pm X^{'-}=\frac{1}{4\alpha' p^{+}}(\Xdot^i\pm X^{'i})^2, \label{eq:virasorolightcone}
\end{align}
where $i=1, \cdots ,d-1$. Inserting the mode expansion (\ref{eq:modeexp}) into (\ref{eq:virasorolightcone}), we find\footnote{The symbol $:~:$ denotes the normal ordering.}
\begin{align}
\alpha_n^{-}=\frac{1}{\sqrt{2\alpha'}p^{+}}\bigg[\frac{1}{2}\sum_{m\in \mathbf{Z}}:\alpha_{n-m}^i\alpha_{mi}:-\delta_{n,0} \bigg], \label{eq:mode-}
\end{align}
where  $\alpha_0^{\mu}\equiv \sqrt{2\alpha'}p^{\mu}$. 
The $n=0$ mode of (\ref{eq:mode-}) gives the mass shell condition,
\begin{align}
M^2=-p^{\mu}p_{\mu}=2p^{+}p^{-}-p^ip_i=\frac{1}{\alpha'}(N-1), \label{eq:massformula}
\end{align}
where $M$ is the mass of the open string and
\begin{align}
N=\sum_{n=1}^{\infty}:\alpha_{-n}^{i}\alpha_{ni}: \label{eq:level}
\end{align}
is the level of the open string. 

\subsection{Observables in  highly excited string states}
Next, we review how to evaluate observables for string states whose level is much larger than $1$ \cite{Damour:1999aw}. At first, we introduce the formal partition function,\footnote{Since $\beta$ is a conjugate parameter of the string level $N$, not a conjugate parameter of the string mass $M$, $\beta$ is not the inverse temperature.}
\begin{align}
Z(\beta)&=tr (e^{-\beta N}) \notag \\
&= \sum_{\{N_n^i\}}\langle \{N_n^i\}|\exp (-\beta N)|\{N_n^i\}\rangle \notag \\
&=\sum_{\{N_n^i\}}\exp (-\beta N[N_n^i]), \label{eq:partitionfunc}
\end{align}
where 
\begin{align}
N[N_n^i]&=\sum_{n=1}^{\infty}\sum_{i=1}^{d-1}n N_n^i, \\
N_n^i&= a_n^{i\dagger}a_n^i,
\end{align}
and  
\begin{align}
a_n^i=\frac{1}{\sqrt{n}}\alpha_n^i, ~~~a_n^{i\dagger}=\frac{1}{\sqrt{n}}\alpha_{-n}^i
\end{align}
with positive $n$. The oscillators $a_n^i$ satisfy the following commutation relations:
\begin{align}
[a_m^i, a_n^{j\dagger}]=\delta^{ij}\delta_{mn}.
\end{align}
The summations in (\ref{eq:partitionfunc}) are taken over all occupation numbers $N_n^i=0,1,2,\cdots$.
 
The density matrix is defined by
\begin{align}
\rho=\frac{\exp (-\beta N)}{Z}.
\end{align}
Thus, the mean value of the level $\Nbar$ and the fluctuation of the level around the mean value $\Delta N$ are given by
\begin{align}
\bar{N}&=\langle N \rangle_{\beta}=-\frac{\del \ln Z}{\del \beta}, \\
(\Delta N)^2&=\langle (N-\Nbar)^2 \rangle_{\beta}=\frac{\del^2 \ln Z}{\del \beta^2},
\end{align}
where $\langle A \rangle_{\beta}\equiv tr (\rho A)$. Since 
\begin{align}
\ln Z&=-(d-1)\sum_{n=1}^{\infty} \ln (1-e^{-\beta n}) \notag  \\
&\stackrel{\beta\ll 1}{\sim}-(d-1)\int_0^{\infty}\frac{dx}{\beta} \ln (1-e^{-x}) \notag \\
&=\frac{(d-1)\pi^2}{6\beta}, \label{eq:lnz}
\end{align}
we obtain 
\begin{align}
\Nbar&=\frac{(d-1)\pi^2}{6\beta^2}, \label{eq:levelvalue} \\
(\Delta N)^2&=\frac{(d-1)\pi^2}{3\beta^3}.
\end{align}
Thus,
\begin{align}
\frac{(\Delta N)^2}{\Nbar^2}=\frac{12\beta}{(d-1)\pi^2}. \label{eq:ratio}
\end{align}
To make the ratio of the fluctuation to the mean value (\ref{eq:ratio}) be small, we impose $\beta\ll 1$ as we have assumed in (\ref{eq:lnz}). With $\beta \ll 1$, $\Nbar$ is much larger than $1$. Thus, we can approximately obtain an observable for string states with $N\gg 1$ by evaluating $\langle A \rangle_{\beta}$ with $\beta\ll 1$, where $A$ is an observable.

Since the mass of the string is given by (\ref{eq:massformula}), we find
\begin{align}
M^2\simeq \frac{\Nbar}{\alpha'}=\frac{(d-1)\pi^2}{6\beta^2 \alpha'}. \label{eq:massbeta}
\end{align}
The entropy of the string, which is the logarithm of the number of string configurations, is found to be
\begin{align}
S=-tr(\rho \ln \rho)=\ln Z +\beta \Nbar=2\pi\sqrt{\frac{(d-1) \Nbar}{6}}. \label{eq:entropy}
\end{align}
This result is consistent with the Cardy formula \cite{Cardy:1986ie}:
\begin{align}
S=2\pi\sqrt{\frac{c N}{6}},
\end{align}
where $c$ is the central charge. In fact, since there are $d-1$ bosonic oscillators, the central charge is
\begin{align}
c=d-1. \label{eq:centralcharge}
\end{align}

\section{Shear viscosity of a highly excited string} \label{sec:viscosstr}
\subsection{Stress tensor of the open string} \label{sec:stress}
Let us find the expression for the stress tensor of the open string. Since (\ref{eq:s1}) can be written as
\begin{align}
S_1=\frac{1}{4\pi \alpha'} \int d^{d+1}x\int d^2 \sigma ~(\Xdot^{\mu}\Xdot^{\nu}-X^{'\mu}X^{'\nu})\delta^{d+1}(x-X)h_{\mu\nu}(x), \label{eq:s1th}
\end{align}
the stress tensor of the open string is given by
\begin{align}
T^{\mu\nu}(x^{+}, x^{-}, x^i)=\frac{1}{4\pi \alpha^{'2}p^{+}} \int_0^{\pi} d\sigma (\dot{X}^{\mu}\dot{X}^{\nu}-X^{'\mu}X^{'\nu})\delta(x^{-}-X^{-}) \delta^{d-1}(x^i-X^i), \label{eq:stress}
\end{align}
where we have used (\ref{eq:lightconegauge}) and $\tau$ is identified with $x^+$ as follows:
\begin{align}
\tau=\frac{x^{+}}{2\alpha' p^{+}}.
\end{align}
Since $x^{+}$ is proportional to $\tau$, $x^{+}$ plays the role of time and $x^{-}$ is the longitudinal spatial coordinate \cite{Green:1987sp}.

We set $\bar{x}^{\mu}=p^i=0$ for simplicity and denote $L_{-}$ and $L$ as the root mean square sizes of the open string for the directions of $x^{-}$ and $x^i$, respectively. We restrict the ranges of the spatial coordinates $x^{-}$ and $x^{i}$ to
\begin{align}
R_d : ~-L_{-}\leq x^{-}-X^-_{cm} \leq L_{-}, ~~~-L\leq x^{i}\leq L, \label{eq:rangex}
\end{align}
because the stress tensor (\ref{eq:stress}) vanishes outside the region $R_d$. The typical sizes of the free open string are \cite{Damour:1999aw, Salomonson:1985eq, Mitchell:1987hr, Mitchell:1987th}
\begin{align}
L&\sim l_s\sqrt{l_sM}, \label{eq:size} \\
L_{-}&\sim \frac{(l_sM)^{3/2}}{p^{+}}. \label{eq:sizeminus}
\end{align}
They are shown in the Appendix \ref{sec:calsizes}.

We consider the zero mode of the stress tensor for the spatial directions. Since the delta functions in (\ref{eq:stress}) are expanded as
\begin{align}
\delta(x^{-}-X^{-})&=\frac{1}{2L_{-}}\sum_{n\in \mathbf{Z}}\exp \bigg(i\frac{\pi n}{L_{-}}(x^{-}-X^{-})\bigg), \label{eq:lminusexp} \\
\delta(x^{i}-X^{i})&=\frac{1}{2L}\sum_{n\in \mathbf{Z}}\exp \bigg(i\frac{\pi n}{L}(x^{i}-X^{i})\bigg),
\end{align}
the zero mode of the stress tensor is
\begin{align}
\Tbar^{\mu\nu}(x^{+})=\frac{1}{4\pi \alpha^{'2}p^{+}V_d} \int_0^{\pi} d\sigma (\dot{X}^{\mu}\dot{X}^{\nu}-X^{'\mu}X^{'\nu}), \label{eq:stress1}
\end{align}
where $V_d=2^dL_{-}L^{d-1}$. This expression is similar to the stress tensor of a polymer appearing in polymer physics \cite{Doi:1986}. 
 
We assume that the nonvanishing component of the metric perturbation is $h_{ij}$ and it depends only on $x^{+}$.
Then, (\ref{eq:s1th}) becomes
\begin{align}
S_1=\frac{V_d}{2}\int dx^{+} ~\Tbar^{ij}(x^{+})h_{ij}(x^{+}).
\end{align}
This is the source term to obtain the shear viscosity of the open string.

\subsection{Linear response and Kubo's formula} \label{sec:visco}
We briefly summarize the response function and Kubo's formula \cite{Chaikin:book, Son:2007vk, Gubser:2008sz}. The deviation of the expectation value of $\Tbar_{ij}(x^{+})$ from its equilibrium value to the first order change in $h^{ij}(x^{+})$ is given by 
\begin{align}
\langle \Tbar_{ij}(x^{+}) \rangle_{\beta, h}=\int_{-\infty}^{\infty}dx^{'+} \chitilde_{ij,kl}(x^{+}-x^{'+})h^{kl}(x^{'+}), \label{eq:linearresponse}
\end{align}
where $\chitilde_{ij,kl}(x^{+}-x^{'+})$ is the response function,
\begin{align}
\chitilde_{ij,kl}(x^{+}-x^{'+})&=2i\theta(x^{+}-x^{'+})\chitilde_{ij,kl}''(x^{+}-x^{'+}), \\
\chitilde_{ij,kl}''(x^{+}-x^{'+})&=\frac{V_d}{4}\langle :[\Tbar_{ij}(x^{+}), \Tbar_{kl}(x^{'+})]:\rangle_{\beta}, \label{eq:comtt}
\end{align}
and 
\begin{align}
\theta(x^{+}-x^{'+})=\lim_{\epsilon \to 0}\int_{-\infty}^{\infty}\frac{d\omega}{2\pi i}\frac{e^{i\omega (x^{+}-x^{'+})}}{\omega -i\epsilon} \label{eq:step}
\end{align}
is a step function \cite{Chaikin:book}. We have introduced the normal ordering in (\ref{eq:comtt}) to avoid the divergence of the response function which comes from the zero point energy \cite{Damour:1999aw, Susskind:1993aa, Mezhlumian:1994pe, Larsen:1998sh}.

Let us define the Laplace transformation of $\chitilde_{ij,kl}(x^{+})$,
\begin{align}
\chi_{ij,kl}(z)=\int_{0}^{\infty} dx^{+} \chitilde_{ij,kl}(x^{+})e^{izx^{+}},
\end{align}
where $z$ is a complex number.  It is shown that $\chi_{ij,kl}(z)$ is obtained from the Fourier transformation of $\chitilde_{ij,kl}''(x^{+})$ as follows \cite{Chaikin:book}:
\begin{align}
\chi_{ij,kl}(z)=\int_{-\infty}^{\infty}\frac{d\omega}{\pi}\frac{\chi''_{ij,kl}(\omega)}{\omega-z}, \label{eq:responsez}
\end{align}
where
\begin{align}
\chi''_{ij,kl}(\omega)=\int_{-\infty}^{\infty} dx^{+} \chitilde''_{ij,kl}(x^{+}) e^{i\omega x^{+}}.
\end{align}
If we set $z=\omega+i\epsilon~(\epsilon\ll 1)$, (\ref{eq:responsez}) becomes
\begin{align}
\chi_{ij,kl}(\omega+i\epsilon)=\int_{-\infty}^{\infty}\frac{d\omega'}{\pi}\frac{\chi''_{ij,kl}(\omega')}{\omega'-\omega-i\epsilon}.
\end{align}
Using the identity
\begin{align}
\frac{1}{\omega'-\omega-i\epsilon}=\mathcal{P}\frac{1}{\omega'-\omega}+i\pi\delta(\omega'-\omega),
\end{align}
where $\mathcal{P}$ denotes the principal value, we obtain
\begin{align}
\chi_{ij,kl}(\omega)=\chi_{ij,kl}'(\omega)+i\chi_{ij,kl}''(\omega), \label{eq:relaimag}
\end{align}
where
\begin{align}
\chi_{ij,kl}'(\omega)=\mathcal{P}\int_{-\infty}^{\infty}\frac{d\omega'}{\pi}\frac{\chi''_{ij,kl}(\omega')}{\omega'-\omega}. \label{eq:realchi}
\end{align}
Thus,  $\chi'_{ij,kl}(\omega)$ and $\chi_{ij,kl}''(\omega)$ are the real part and the imaginary part of $\chi_{ij,kl}(\omega)$, respectively.
Carrying out the Fourier transformation of (\ref{eq:linearresponse}), we obtain 
\begin{align}
\langle \Tbar_{ij}(\omega)\rangle_{\beta, h}=\chi_{ij,kl}(\omega)h^{kl}(\omega). \label{eq:linearres}
\end{align}
Comparing (\ref{eq:linearres}) with the stress tensor of a viscous fluid in a curved background \cite{Son:2007vk, Gubser:2008sz}, we obtain the Kubo's formula for  shear viscosity,\footnote{The factor 2 in (\ref{eq:kubo}) comes from the symmetric property of the indices $k$ and $l$ in (\ref{eq:linearres}).}
\begin{align}
\eta=\lim_{\omega\to 0}\frac{2\chi''_{12,12}(\omega)}{\omega}. \label{eq:kubo}
\end{align}

\subsection{Calculation of the shear viscosity of the string} 
Let us calculate the response function.  Using the mode expansion (\ref{eq:modeexp}), the stress tensor (\ref{eq:stress1}) becomes
\begin{align}
\Tbar_{ij}(x^{+})=\frac{1}{2\alpha' p^{+}V_d}\sum_{\stackrel {\scriptstyle n\neq 0}{\scriptstyle n\in \mathbf{Z}}}\alpha_n^i\alpha_n^je^{-2in\tau}. \label{eq:stressoscillator}
\end{align}
Inserting (\ref{eq:stressoscillator}) into  (\ref{eq:comtt}), we find
\begin{align}
\chi''_{ij,kl}(x^{+}-x^{'+})=\frac{1}{16\alpha^{'2}p^{+2}V_d}\sum_{\stackrel {\scriptstyle n\neq 0}{\scriptstyle n\in \mathbf{Z}}}\sum_{\stackrel {\scriptstyle m\neq 0}{\scriptstyle m\in \mathbf{Z}}}e^{-2in\tau}e^{-2im\tau'}\langle :[\alpha_n^i\alpha_n^j, \alpha_m^k\alpha_m^l]: \rangle_{\beta}. \label{eq:res1}
\end{align}
Using the identity
\begin{align}
[AB,CD]=[A,C]BD+A[B,C]D+C[A,D]B+CA[B,D],
\end{align}
we obtain
\begin{align}
\langle :[\alpha_n^i\alpha_n^j, \alpha_m^k\alpha_m^l]: \rangle_{\beta}=n\delta_{n+m,0}&[\delta^{ik}\langle :\alpha_n^j\alpha_m^l : \rangle_{\beta}+\delta^{jk}\langle :\alpha_n^i\alpha_m^l :\rangle_{\beta} \notag \\
&+\delta^{il}\langle :\alpha_m^k\alpha_n^j :\rangle_{\beta}+\delta^{jl}\langle :\alpha_m^k\alpha_n^i :\rangle_{\beta}]. \label{eq:comresult}
\end{align}
Inserting (\ref{eq:comresult}) into (\ref{eq:res1}),  the first term becomes
\begin{align}
&~~~\sum_{\stackrel {\scriptstyle n\neq 0}{\scriptstyle n\in \mathbf{Z}}}\sum_{\stackrel {\scriptstyle m\neq 0}{\scriptstyle m\in \mathbf{Z}}}e^{-2in\tau}e^{-2im\tau'} n\delta_{n+m,0}\delta^{ik}\langle :\alpha_n^j\alpha_m^l : \rangle_{\beta} \notag \\
&=\sum_{\stackrel {\scriptstyle n\neq 0}{\scriptstyle n\in \mathbf{Z}}} e^{-2in(\tau-\tau')}n\delta^{ik}\langle :\alpha_n^j\alpha_{-n}^l : \rangle_{\beta} \notag \\
&=\sum_{n=1}^{\infty}n^2 \delta^{ik}(e^{-2in(\tau-\tau')}\langle :a_n^{l\dagger}a_n^j : \rangle_{\beta}-e^{2in(\tau-\tau')}\langle :a_n^{j\dagger}a_n^l : \rangle_{\beta}). \label{eq:halfway}
\end{align}
Using the formula \cite{Damour:1999aw}
\begin{align}
\langle :a_n^{i\dagger}a_m^j : \rangle_{\beta}=\frac{\delta_{nm}\delta^{ij}}{e^{\beta n}-1}, \label{eq:ensumble}
\end{align}
(\ref{eq:halfway}) becomes
\begin{align}
&\sum_{\stackrel {\scriptstyle n\neq 0}{\scriptstyle n\in \mathbf{Z}}}\sum_{\stackrel {\scriptstyle m\neq 0}{\scriptstyle m\in \mathbf{Z}}}e^{-2in\tau}e^{-2im\tau'} n\delta_{n+m,0}\delta^{ik}\langle :\alpha_n^j\alpha_m^l : \rangle_{\beta} \notag \\
&=-2i\sum_{n=1}^{\infty}\frac{n^2}{e^{\beta n}-1}\delta^{ik}\delta^{jl}\sin (2n (\tau-\tau')).
\end{align}
Calculating the remaining terms, we find
\begin{align}
\chitilde_{ij,kl}''(x^{+}-x^{'+})=\frac{-i}{4\alpha^{'2} p^{+2}V_d}\delta_{ij,kl}\sum_{n=1}^{\infty}\frac{n^2}{e^{\beta n}-1} \sin \bigg(\frac{n}{\alpha'p^{+}}(x^{+}-x^{'+}) \bigg),
\end{align}
where $\delta_{ij,kl}\equiv \delta_{ik}\delta_{jl}+\delta_{il}\delta_{jk}$.
 
Since
\begin{align}
\int_{-\infty}^{\infty} d(x^{+}-x^{'+}) e^{i\omega (x^{+}-x^{'+})}\sin \bigg(\frac{n}{\alpha'p^{+}}(x^{+}-x^{'+}) \bigg)=\frac{\pi}{i}\bigg(\delta(\omega+\frac{n}{\alpha'p^{+}})-\delta(\omega-\frac{n}{\alpha'p^{+}})\bigg),
\end{align}
we obtain
\begin{align}
\chi''_{ij,kl}(\omega)=\frac{\pi}{4\alpha^{'2} p^{+2}V_d}\delta_{ij,kl}\sum_{n=1}^{\infty}\frac{n^2}{e^{\beta n}-1}\bigg(\delta(\omega-\frac{n}{\alpha'p^{+}})-\delta(\omega+\frac{n}{\alpha'p^{+}})\bigg). \label{eq:imchipp}
\end{align}
If we assume that $\omega$ is positive, 
\begin{align}
\chi''_{ij,kl}(\omega)=\frac{\alpha' p^{+}\pi}{4V_d} \frac{\omega^2}{e^{\beta \alpha' p^{+}\omega}-1}\delta_{ij,kl}. 
\end{align}
Taking the low frequency limit, it becomes
\begin{align}
\chi''_{ij,kl}(\omega)\stackrel{\omega\to 0}{\sim} \frac{M}{4V_d}\sqrt{\frac{6\alpha'}{d-1}}\delta_{ij,kl}~\omega,
\end{align}
where we have used (\ref{eq:massbeta}).
From (\ref{eq:kubo}), we obtain the shear viscosity of the highly excited open string as follows:
\begin{align}
\eta=\sqrt{\frac{6}{d-1}}\frac{Ml_s}{2V_d}. \label{eq:shear} 
\end{align}
Since from (\ref{eq:entropy}), the entropy density $s$ is given by
\begin{align} 
s=\frac{S}{V_d}=2\pi \sqrt{\frac{d-1}{6}}\frac{Ml_s}{V_d}, \label{eq:entropydes}
\end{align}
we obtain the ratio of shear viscosity to entropy density,
\begin{align}
\frac{\eta}{s}=\frac{3}{2 \pi (d-1)}=\frac{3}{2 \pi c}, \label{eq:etaovers}
\end{align}
where we have used (\ref{eq:centralcharge}).

\section{Shear viscosity of a string on a stretched horizon and black hole membrane paradigm} \label{sec:comparison}
\subsection{Stretched horizon of Schwarzshild black hole}
Let us define the stretched horizon of the Schwarzshild black hole \cite{Susskind:1993ws}.
The metric of the $(d+1)$-dimensional Schwarzshild black hole is given by
\begin{align}
ds^2=-h(r)dt^2+\frac{1}{h(r)}dr^2+r^2d\Omega_{d-1}^2, \label{eq:schwartzshild}
\end{align}
where
\begin{align}
h(r)&=1-\bigg(\frac{r_{H}}{r}\bigg)^{d-2}, \\
r_{H}^{d-2}&=\frac{16\pi MG}{(d-1)\Omega_{d-1}}, \label{eq:horizonrad}
\end{align}
and $\Omega_{d-1}=2\pi^{d/2}/\Gamma (d/2)$ is the volume of a unit $(d-1)$-dimensional sphere. Here, $r_{H}$ is the radius of the event horizon, $M$ is the Arnowitt-Deser-Misner (ADM) mass of the black hole  and $G$ is the Newton's constant. The relation between $G$ and the string coupling constant $g_s$ is given by \cite{Green:1987sp}
\begin{align}
G\sim g_s^2l_s^{d-1}. \label{eq:ggs}
\end{align}

To obtain the near horizon geometry of the black hole, we set \cite{Townsend:1997ku}
\begin{align}
r^{d-2}-r_{H}^{d-2}=\frac{(d-2)^2}{4}r_{H}^{d-4}\rho^2. \label{eq:defofx}
\end{align}
Then, for $r\sim r_{H}$, the metric (\ref{eq:schwartzshild}) becomes the Rindler metric,
\begin{align}
ds^2\sim -\kappa^2 \rho^2 dt^2+d\rho^2+r_{H}^2d\Omega_{d-1}^2, \label{eq:rindler}
\end{align}
where 
\begin{align} 
\kappa =\frac{d-2}{2r_H}
\end{align} 
is the surface gravity.  This metric is valid for 
\begin{align}
0\leq \rho\ll \frac{1}{\kappa}~,
\end{align}
and the event horizon is located at $\rho=0$. By a periodic identification of the Euclidean time coordinate \cite{Townsend:1997ku}, we obtain the Hawking temperature,
\begin{align}
T_{H}=\frac{\kappa}{2\pi}.
\end{align}
By using the first law of thermodynamics, $dM=TdS$, we obtain the Bekenstein-Hawking entropy,
\begin{align}
S_{BH}=\frac{A_{d-1}}{4G}, \label{eq:bhentropy}
\end{align}
where 
\begin{align}
A_{d-1}=\Omega_{d-1}r_H^{d-1}\sim (l_s M)^{\frac{d-1}{d-2}}g_s^{\frac{2}{d-2}}
\end{align}
is the area of the event horizon.
 
To find the energy and temperature for an observer at a radial coordinate $\rho$, it is convenient to introduce dimensionless Rindler quantities \cite{Susskind:1993ws, Halyo:1996vi, Halyo:1996ub, Halyo:2001us}. At first, we define the Rindler time as
\begin{align}
\theta\equiv \kappa t.
\end{align}
Since Rindler energy $E_R$ is conjugate to $\theta$ and $M$ is conjugate to $t$, we find
\begin{align}
[E_{R}, \theta ]&=i, \label{eq:erthetacom} \\
[M, t]&=i. \label{eq:mtcom} 
\end{align}
From (\ref{eq:mtcom}),  $t$ can be expressed by $-i\del/\del M$. Thus, (\ref{eq:erthetacom}) is expressed as
\begin{align}
\frac{\del E_{R}}{\del M}=\frac{1}{\kappa}. \label{eq:delexpofer}
\end{align}
Integrating (\ref{eq:delexpofer}) over $M$, we obtain the Rindler energy,
\begin{align}
E_R=\frac{A_{d-1}}{8\pi G}.
\end{align}
Using the first law of thermodynamics, we find the Rindler temperature,
\begin{align}
T_R=\frac{1}{2\pi}.
\end{align}
Since the proper time at the  radial coordinate $\rho$ is
\begin{align}
\tau|_{\rho}=\rho \theta, 
\end{align}
the energy and temperature for an observer at $\rho$ become
\begin{align}
E_{\rho}&=\frac{E_{R}}{\rho}=\frac{A_{d-1}}{8\pi G \rho},  \\
T_{\rho}&=\frac{T_{R}}{\rho}=\frac{1}{2\pi \rho}. \label{eq:xtemp}
\end{align}
The stretched horizon is defined by the place where the local Unruh temperature is given by the Hagedorn temperature $T\sim 1/l_s$ \cite{Susskind:1993ws}. Thus, from (\ref{eq:xtemp}), we find that the stretched horizon is located at $\rho\sim l_s$. 
On the stretched horizon, the energy and temperature become \cite{Susskind:1993ws, Halyo:1996vi, Halyo:1996ub, Halyo:2001us}
\begin{align}
E_{sh}&=\frac{E_{R}}{l_s}=\frac{A_{d-1}}{8\pi G l_s}, \label{eq:shmass} \\
T_{sh}&=\frac{T_{R}}{l_s}=\frac{1}{2\pi l_s}. \label{eq:shtemp}
\end{align}

The difference between $E_{sh}$ and $M$ is caused by the red shift of clock rates that takes place between the stretched horizon and the asymptotic infinity. This implies that $E_{sh}$ is renormalized to $M$ by the classical gravitational field outside the stretched horizon \cite{Susskind:1993ws}. 

If the red shift factor is of the order of $1$, namely,
\begin{align}
\frac{\tau|_{r=\infty}}{\tau|_{\rho\sim l_s}}=\frac{t}{\kappa l_s t}=\frac{1}{\kappa l_s}\sim \mathcal{O}(1),
\end{align}
we find \cite{Susskind:1993ws}
\begin{align}
r_H&\sim l_s, \\
g_s^2&\sim \frac{1}{l_s M}. \label{eq:criticalstringcoupling}
\end{align}
The expression (\ref{eq:criticalstringcoupling}) is the same as the critical string coupling in the string-black hole correspondence \cite{Horowitz:1996nw}. In this situation, the energy and temperature for an observer on the stretched horizon are of the same order as those for the asymptotic observer.

If we assume that a highly excited string is on the stretched horizon, the Bekenstein-Hawking entropy (\ref{eq:bhentropy}) can be reproduced by the entropy of the string except for a numerical coefficient \cite{Susskind:1993ws}. Since the string possesses the energy $E_{sh}$, the level of the string is 
\begin{align}
N=l_s^2 E_{sh}^2.
\end{align}
Thus, the entropy of the string is 
\begin{align}
S\sim \sqrt{N}\sim \frac{A_{d-1}}{G},
\end{align}
which is of the same order as the Bekenstein-Hawking entropy (\ref{eq:bhentropy}).

\subsection{Shear viscosity of the string on the stretched horizon}
Let us consider the shear viscosity of the string on the stretched horizon. In the membrane paradigm, no radial thickness of the fictitious membrane is considered. Therefore, the mass dimension of the stress tensor of the membrane is given by $d$, which is one less than that of (\ref{eq:stress1}). Thus, we would like to forget the longitudinal thickness of the string. To do so, we define the longitudinally reduced stress tensor of the string. This is defined by the zero mode of the stress tensor for the longitudinal direction. Since (\ref{eq:stress}) can be expanded as
\begin{align}
T^{ij}(x^+, x^-, x^i)=\frac{1}{2L_{-}}\sum_{n\in \mathbf{Z}}\tilde{T}^{ij}(x^+, n, x^i)e^{i\frac{\pi n}{L_{-}}x^{-}},
\end{align}
the longitudinally reduced stress tensor is  $\tilde{T}^{ij}(x^+, 0, x^i)\equiv T_r^{ij}(x^+, x^i)$.
From (\ref{eq:stress}) and (\ref{eq:lminusexp}), we find
\begin{align}
T_r^{ij}(x^{+},x^i)=\frac{1}{4\pi \alpha^{'2}p^{+}} \int_0^{\pi} d\sigma (\dot{X}^{i}\dot{X}^{j}-X^{'i}X^{'j})\delta^{d-1}(x^k-X^k).
\end{align}
We see that the mass dimension of the reduced stress tensor is $d$, which is one less than that of the original stress tensor. Since the zero mode of $T_r^{ij}(x^{+}, x^i)$ for the transverse directions becomes
\begin{align}
\Tbar_r^{ij}(x^{+})&=\frac{1}{4\pi \alpha^{'2}p^{+}(2L)^{d-1}} \int_0^{\pi} d\sigma (\dot{X}^{i}\dot{X}^{j}-X^{'i}X^{'j}) \notag \\
&=2L_{-}\Tbar^{ij}(x^{+}),
\end{align}
the shear viscosity of the longitudinally reduced string is 
\begin{align}
\eta_r=\sqrt{\frac{6}{d-1}}\frac{Ml_s}{2V_{d-1}}, \label{eq:reducedeta}
\end{align}
where $V_{d-1}\equiv (2L)^{d-1}$. The ratio of shear viscosity to entropy density (\ref{eq:etaovers}) does not change even if the string is longitudinally reduced because the volume factors in (\ref{eq:shear}) and (\ref{eq:entropydes}) have been canceled.

Let us estimate the shear viscosity of the longitudinally reduced string on the stretched horizon by using (\ref{eq:reducedeta}). 
Since $V_{d-1}$ in (\ref{eq:reducedeta}) is replaced by the area of the stretched horizon and the mass of the string on the stretched horizon is given by (\ref{eq:shmass}), we find that the shear viscosity of the string on the stretched horizon is 
\begin{align}
\eta_r^{sh}\sim \frac{E_{sh}l_s}{r_{H}^{d-1}}\sim \frac{1}{G}.
\end{align}
Surprisingly, this is of the same order as the shear viscosity of the fictitious membrane in the membrane paradigm, which is known as \cite{Thorne:1986iy}
\begin{align}
\eta_{BH}=\frac{1}{16\pi G}.
\end{align}

This correspondence can also be derived along with the string-black hole correspondence \cite{Horowitz:1996nw}. At the critical string coupling $g_s^2\sim 1/l_sM$, a highly excited string becomes a black hole with the radius $r_H\sim l_s$ \cite{Damour:1999aw, Horowitz:1997jc, Chialva:2009pf}. Thus, at this point, the shear viscosity of the string becomes
\begin{align}
\eta^{sh}_r\sim \frac{l_sM}{l_s^{d-1}}.
\end{align}
On the other hand, the shear viscosity in the membrane paradigm becomes
\begin{align}
\eta_{BH}\sim \frac{1}{g_s^2l_s^{d-1}}\sim \frac{l_sM}{l_s^{d-1}},
\end{align}
where we have used (\ref{eq:ggs}). Thus, we find that the shear viscosity of the string is consistent with the membrane paradigm at the critical string coupling.

In our estimate, $\eta/s$ does not change even if we put the string on the stretched horizon. On the other hand, $\eta/s$ in the membrane paradigm is known as \cite{Thorne:1986iy}
\begin{align}
\frac{\eta_{BH}}{s_{BH}}=\frac{1}{4\pi}.
\end{align}
Thus, if the central charge is $c=6$, $\eta/s$ of the string matches with that of the membrane paradigm. This value of the central charge has been discussed to reproduce the correct numerical coefficient of the Bekenstein-Hawking entropy of the Schwarzshild black hole from string theory  \cite{Halyo:1996xe}.

\section{Summary and comments} \label{sec:summary}
We have obtained the shear viscosity and the ratio of shear viscosity to entropy density of the highly excited string by using the Kubo's formula. The results have been given by (\ref{eq:shear}) and (\ref{eq:etaovers}). Then, we have discussed the shear viscosity of the highly excited string on the stretched horizon of the Schwarzshild black hole. In order to compare the obtained results with the results of the membrane paradigm, we have  reduced the longitudinal thickness of the string. 
We have found that the shear viscosity of the highly excited string on the stretched horizon matches with that in the membrane paradigm except for a numerical coefficient. Also, we have found that the value of $\eta/s$ of the string agrees with that in the membrane paradigm when the central charge is $6$, which has been discussed to reproduce the correct numerical coefficient of the Bekenstein-Hawking entropy of the Schwarzshild black hole from string theory in  \cite{Halyo:1996xe}.

Five comments are in order:
\begin{itemize}
\item We have not considered the self-interactions of the highly excited string although there are several discussions about the self-interactions of the highly excited string \cite{Damour:1999aw, Horowitz:1997jc, Chialva:2009pf}. The self-interactions of the highly excited string will lead to  the $g_s$ corrections to the shear viscosity of the string. It might be interesting to see whether $\eta/s$ of the highly excited string is modified by the $g_s$ corrections. 
\item It is important to investigate whether the correct numerical coefficient of the shear viscosity in the membrane paradigm can be derived from superstring theory. 
\item We have not discussed the bulk viscosity because we could not reproduce the negative bulk viscosity of the membrane paradigm from the highly excited string on the stretched horizon. 
\item It would be interesting to find transport coefficients of a highly excited string when source fields are given by other fields instead of metric.
\item In Sec. \ref{sec:comparison}, we have put the string on the stretched horizon. Since the temperature of the stretched horizon is given by the Hagedorn temperature, we will be able to  interpret the shear viscosity of the string on the stretched horizon as the shear viscosity of the string at the Hagedorn temperature \cite{Susskind:1993ws, Susskind:1993aa, Mezhlumian:1994pe, Larsen:1998sh}. Thus, we expect that if we put the string at a place whose radial coordinate is labeled by $\rho$, we can obtain the shear viscosity of the string at a temperature given by (\ref{eq:xtemp}). 
\end{itemize}
These considerations should be investigated in  future works.

\section*{Acknowledgments}
We would like to thank to Masud Chaichian, Shinsuke Kawai, Esko Keski-Vakkuri, Hiroto Ogawa, Anca Tureanu and Patta Yogendran for useful discussions and comments. 
Y.S. was supported in part by JSPS-Academy of Finland bilateral scientist exchange programme.

\appendix
\section{Typical sizes of free open string} \label{sec:calsizes}
We show that the typical sizes of the highly excited free open string are given by (\ref{eq:size}) and (\ref{eq:sizeminus}) \cite{Damour:1999aw, Salomonson:1985eq, Mitchell:1987hr, Mitchell:1987th}. At first, we calculate the transverse size of the open string, which is defined by \cite{Damour:1999aw}:
\begin{align}
L^2\equiv \lim_{T\to \infty} \int_{-T}^{T}~\frac{d\tau}{2T}\int_0^{\pi} \frac{d\sigma}{\pi}  ~ \langle :(X^i-\bar{x}^i)^2 :\rangle_{\beta},
\end{align}
where we have assumed $p^i=0$. Using the mode expansion (\ref{eq:modeexp}), we find
\begin{align}
L^2&=2\alpha' \sum_{n=1}^{\infty} \frac{1}{n} \langle : a_n^{i \dagger}a_{ni} : \rangle_{\beta} \notag \\
&=2(d-1)\alpha' \sum_{n=1}^{\infty} \frac{1}{n} \frac{1}{e^{\beta n}-1},
\end{align}
where we have used (\ref{eq:ensumble}). Let us evaluate the following sum:
\begin{align}
I_1(\beta)\equiv \sum_{n=1}^{\infty} \frac{1}{n} \frac{1}{e^{\beta n}-1}.
\end{align}
Since $\beta \ll 1$, we can approximate the summation by an integration  as follows:
\begin{align}
I_1(\beta)\sim \int_{\beta}^{\infty} dx \frac{1}{x} \frac{1}{e^{x}-1}, \label{eq:ccc}
\end{align}
where we have set $x\equiv \beta n$. Since 
\begin{align}
\frac{d I_1(\beta)}{d \beta}&=-\frac{1}{\beta} \frac{1}{e^{\beta}-1} \notag \\
&\sim -\frac{1}{\beta^2},
\end{align}
we find
\begin{align}
I_1(\beta)\sim \frac{1}{\beta}+C_1,
\end{align}
where $C_1$ is a constant of integration. Since $C_1$ is found to be a constant of the order of $1$ by a numerical calculation, we neglect $C_1$. Thus, we find
\begin{align}
L\sim \sqrt{\frac{2(d-1)\alpha'}{\beta}}\sim l_s\sqrt{l_sM}, \label{eq:eee}
\end{align}
where we have used (\ref{eq:massbeta}).

Next, we calculate the longitudinal size of the string. As in the case of the transverse size of the string, the longitudinal size is defined by
\begin{align}
L_{-}^2\equiv \lim_{T\to \infty} \int_{-T}^{T}~\frac{d\tau}{2T}\int_0^{\pi} \frac{d\sigma}{\pi}  ~ \langle :(X^{-}-X_{cm}^{-})^2 :\rangle_{\beta},
\end{align}
where $X_{cm}^{-}\equiv \bar{x}^{-}+2\alpha'p^{-}\tau$ is the center-of-mass position of the string in the longitudinal direction. Using the mode expansion (\ref{eq:modeexp}), we find
\begin{align}
L_{-}^2=2\alpha' \sum_{n=1}^{\infty}\frac{\langle : \alpha_{-n}^{-}\alpha_n^{-} : \rangle_{\beta}}{n^2}.
\end{align}
Now, the mode $\alpha_n^{-}$ is given by (\ref{eq:mode-}). Rewriting (\ref{eq:mode-}) in terms of the standard harmonic oscillators $a_n^{i}$, we find
\begin{align}
\alpha_{n}^{-}&=\frac{1}{2\sqrt{2\alpha'}p^{+}}\bigg[2\sum_{m=1}^{\infty}\sqrt{m(n+m)}a_m^{i\dagger}a_{n+m,i}+(1-\delta_{n,1})\sum_{m=1}^{n-1}\sqrt{m(n-m)}a_{n-m}^ia_{mi} \bigg] \notag \\
\alpha_{-n}^{-}&=\frac{1}{2\sqrt{2\alpha'}p^{+}}\bigg[2\sum_{m=1}^{\infty}\sqrt{m(n+m)}a_{m+n}^{i\dagger}a_{mi}+(1-\delta_{n,1})\sum_{m=1}^{n-1}\sqrt{m(n-m)}a_{m}^{i\dagger}a_{n-m,i}^{\dagger} \bigg]
\end{align}
where $n$ is positive. Thus,
\begin{align}
\langle : \alpha_{-n}^{-}\alpha_n^{-} : \rangle_{\beta}&=\frac{1}{8\alpha'p^{+2}}\bigg[4\sum_{m'=1}^{\infty}\sum_{m=1}^{\infty}\sqrt{m'(n+m')}\sqrt{m(n+m)}\langle : a_{n+m'}^{j\dagger}a_{m'j}a_m^{i\dagger}a_{n+m,i}: \rangle_{\beta} \notag \\
&+(1-\delta_{n,1})\sum_{m'=1}^{n-1}\sum_{m=1}^{n-1}\sqrt{m'(n-m')}\sqrt{m(n-m)}\langle : a_{m'}^{j\dagger}a_{n-m',j}^{\dagger}a_{n-m}^{i}a_{m,i}: \rangle_{\beta} \bigg] \notag \\
&=\frac{1}{8\alpha'p^{+2}}\bigg[4(d-1)\sum_{m=1}^{\infty}m(n+m)\frac{1}{e^{\beta(n+m)}-1}\frac{1}{e^{\beta m}-1} \notag \\
&+(1-\delta_{n,1})(d-1)\sum_{m=1}^{n-1}m(n-m)\frac{1}{e^{\beta(n-m)}-1}\frac{1}{e^{\beta m}-1}\bigg].
\end{align}
Therefore, we obtain
\begin{align}
L_{-}^2&=\frac{d-1}{4p^{+2}}\bigg[4\sum_{n=1}^{\infty}\sum_{m=1}^{\infty}\frac{m(n+m)}{n^2}\frac{1}{e^{\beta(n+m)}-1}\frac{1}{e^{\beta m}-1} \notag \\
&+\sum_{n=2}^{\infty}\sum_{m=1}^{n-1}\frac{m(n-m)}{n^2}\frac{1}{e^{\beta(n-m)}-1}\frac{1}{e^{\beta m}-1}\bigg]. \label{eq:l-1}
\end{align}
Next, we evaluate the following summations,
\begin{align}
I_2(\beta)&\equiv \sum_{n=1}^{\infty}\sum_{m=1}^{\infty}\frac{m}{n}\frac{1}{e^{\beta(n+m)}-1}\frac{1}{e^{\beta m}-1}, \label{eq:g1} \\
I_3(\beta)&\equiv \sum_{n=1}^{\infty}\sum_{m=1}^{\infty}\frac{m^2}{n^2}\frac{1}{e^{\beta(n+m)}-1}\frac{1}{e^{\beta m}-1}, \label{eq:g2} \\
I_4(\beta)&\equiv \sum_{n=2}^{\infty}\sum_{m=1}^{n-1}\frac{m(n-m)}{n^2}\frac{1}{e^{\beta(n-m)}-1}\frac{1}{e^{\beta m}-1}. \label{eq:g3}
\end{align}
At first, let us evaluate $I_2(\beta)$ and $I_3(\beta)$. Since we can approximate (\ref{eq:g1}) and (\ref{eq:g2}) by integrations, these become
\begin{align}
I_2(\beta)&\sim \frac{1}{\beta^2}\int_{\beta}^{\infty}dx\frac{x}{e^x-1}\int_{\beta}^{\infty}dy \frac{1}{y}\frac{1}{e^{x+y}-1}, \label{eq:g11} \\
I_3(\beta)&\sim \frac{1}{\beta^2}\int_{\beta}^{\infty}dx\frac{x^2}{e^x-1}\int_{\beta}^{\infty}dy \frac{1}{y^2}\frac{1}{e^{x+y}-1}. \label{eq:g21}
\end{align} 
Using\footnote{(\ref{eq:aaa}) and (\ref{eq:bbb}) can be obtained in the same way as (\ref{eq:ccc}).}
\begin{align}
&\int_{\beta}^{\infty}dy \frac{1}{y}\frac{1}{e^{x+y}-1}\sim -\frac{\ln \beta}{e^x-1}, \label{eq:aaa} \\
&\int_{\beta}^{\infty}dy \frac{1}{y^2}\frac{1}{e^{x+y}-1}\sim \frac{1}{\beta}\frac{1}{e^x-1}, \label{eq:bbb}
\end{align}
(\ref{eq:g11}) and (\ref{eq:g21}) become
\begin{align}
I_2(\beta)&\sim -\frac{\ln \beta}{\beta^2}\int_{\beta}^{\infty}dx\frac{x}{(e^x-1)^2} \notag \\ 
&\sim \frac{(\ln \beta)^2}{\beta^2}, \label{eq:i2}
\end{align}
\begin{align}
I_3(\beta)&\sim \frac{1}{\beta^3}\int_{\beta}^{\infty}dx\frac{x^2}{(e^x-1)^2} \notag \\
&\sim \frac{1}{\beta^3}. \label{eq:i3}
\end{align}

Next, we evaluate $I_4(\beta)$. By using the following inequality, 
\begin{align}
\frac{e^x+1}{e^x-1} < 1+\frac{2}{x},
\end{align}
for $x>0$, (\ref{eq:g3}) becomes
\begin{align}
I_4(\beta)&=\sum_{n=2}^{\infty}\frac{1}{e^{\beta n}-1}\sum_{m=1}^{n-1}\frac{m}{n}\bigg(1-\frac{m}{n}\bigg)\frac{e^{\beta m}+1}{e^{\beta m}-1} \notag \\
&< \sum_{n=2}^{\infty}\frac{1}{e^{\beta n}-1}\sum_{m=1}^{n-1}\frac{m}{n}\bigg(1-\frac{m}{n}\bigg)\bigg(1+\frac{2}{\beta m}\bigg) \notag \\
&=\frac{1}{\beta}\bigg[\sum_{n=2}^{\infty}\frac{1}{e^{\beta n}-1}-\sum_{n=2}^{\infty}\frac{1}{n}\frac{1}{e^{\beta n}-1}\bigg]+\frac{1}{6}\bigg[\sum_{n=2}^{\infty}\frac{n}{e^{\beta n}-1}-\sum_{n=2}^{\infty}\frac{1}{n}\frac{1}{e^{\beta n}-1}\bigg]. \label{eq:g31}
\end{align}
Since the summations in (\ref{eq:g31}) are approximately given by
\begin{align}
\sum_{n=2}^{\infty}\frac{1}{e^{\beta n}-1}&\sim \frac{1}{\beta}\int_{2\beta}^{\infty}dx \frac{1}{e^x-1} \notag \\
&=\frac{1}{\beta}(2\beta-\ln (e^{2\beta}-1)) \notag \\
&\sim \frac{1}{\beta}\ln \bigg( \frac{1}{2\beta} \bigg), \\
\sum_{n=2}^{\infty}\frac{1}{n}\frac{1}{e^{\beta n}-1}&\sim \int_{2\beta}^{\infty}dx \frac{1}{x}\frac{1}{e^x-1} \notag \\
&\sim \frac{1}{2\beta}, \\
\sum_{n=2}^{\infty}\frac{n}{e^{\beta n}-1}&\sim \frac{1}{\beta^2}\int_{2\beta}^{\infty}dx \frac{x}{e^x-1} \notag \\
&\sim \frac{\pi^2}{6\beta^2},
\end{align}
the most dominant term in (\ref{eq:g31}) is the first term. Thus,
\begin{align}
I_4(\beta)\lesssim \frac{1}{\beta^2}\ln \bigg( \frac{1}{2\beta}\bigg). \label{eq:i4}
\end{align}
Among (\ref{eq:i2}), (\ref{eq:i3}) and (\ref{eq:i4}), Eq.(\ref{eq:i3}) is the largest for $\beta \ll 1$. Thus, we obtain
\begin{align}
L_{-}\sim \sqrt{\frac{d-1}{p^{+2}\beta^3}}\sim \frac{(l_sM)^{3/2}}{p^{+}}. \label{eq:ddd}
\end{align}

If the string is static, namely $p^+\sim M$, (\ref{eq:ddd}) becomes of the same order as (\ref{eq:eee}).

\end{document}